\documentclass[final,1p,times]{elsarticle}

\usepackage{amssymb}
\usepackage{amsmath}

\usepackage{bm}

\begin{document}

\begin{frontmatter}



\title{Axion effects on the non-radial oscillations of neutron stars} 

\author[label1]{Deepak Kumar}
\affiliation[label1]{organization={Institute of Physics},
            addressline={Sachivalay Road, near Sainik Schook},
            city={Bhubaneswar},
            postcode={751005},
            state={Odisha},
            country={India}}

\author[label2]{Hiranmaya Mishra}
\affiliation[label2]{organization={School of Physical Sciences,National Institute of Science Education and Research},
            addressline={An OCC of Homi Bhabha National Institute},
            city={Jatni},
            postcode={752050},
            state={Odisha},
            country={India}}


\begin{abstract}
The effects of axions on quark matter equation of state (EOS) is studied  within the three flavor Nambu--Jona-Lasinio model and its effects on  on the non-radial  oscillations of neutron stars is investigated. Using such an EOS for quark matter with axions and a EOS for hadronic matter within the  relativistic mean field (RMF) theory, we discuss the hadron-quark phase transition (HQPT) using the Gibbs construction. The EOS so obtained is used to investigate the structure of hybrid neutron star (NS)s. It is found that the presence of axions in the core of compact stars stabilizes  hybrid NSs  in agreement with modern astrophysical constraints. It is further observed that the quadrupolar fundamental modes ($f$-modes) for such hybrid NSs get substantial enhancements both due to a larger quark core in the presence of axions and from the hyperons as compared to a canonical nucleonic neutron stars.
\end{abstract}







\end{frontmatter}



\section{Introduction and formalism}\label{sec:Introduction}
Neutron stars (NS)s, which are the second densest object in the universe after a black hole, could be used to probe the strong interactions at extreme densities ($\sim$ few times nuclear saturation density, $\rho_0$). At such high densities, it is possible to have a hadron-quark phase transition (HQPT) \cite{Kumar:2024abb}. Macroscopic properties of such NS like mass, radius, moment of inertia, tidal deformability depend on the equation of state (EOS) of matter of NS. These properties (mass and radius) can be found by solving the following Tolman-Oppenheimer-Volkoff (TOV) equations
\begin{eqnarray}
\frac{dp}{dr} = -\left(\epsilon +p \right) \frac{m+4 \pi r^3p}{r(r-2m)}, \qquad \frac{dm}{dr} = 4\pi r^2 \epsilon
\end{eqnarray}
with the boundary conditions: $m(0) = p(R) = 0,$ and $p(r=0) = p_0$.  The non-radial oscillations of NS are estimated by solving the following pulsating equations, which are found by linearizing the perturbed Einstein equations given as, with the 'prime' denoting derivative with respect to radial coordinate,\cite{Kumar:2021hzo},
\begin{eqnarray}
Q'-\frac{1}{c_s^2}\left[\omega^2 r^2e^{\Lambda-2\Phi}Z+\Phi' Q\right]+l(l+1)e^\Lambda Z = 0
\end{eqnarray}
\begin{eqnarray}
Z'-2\Phi' Z+e^\Lambda \frac{Q}{r^2}-\frac{\omega_{BV}^2 e^{-2\Phi}}{\Phi'\left(1-\frac{2m}{r}\right)}\left(Z+\Phi'e^{-\Lambda+2\Phi}\frac{Q}{\omega^2r^2}\right)=0
\end{eqnarray}
where $\omega_{BV}$ is the Brunt-Vaisala frequencies which depends on both equilibrium and adiabatic sound speeds in the matter and $\omega$ is the frequency of the mode, $f$ in the present case. The boundary conditions are as follows: $Q(r) = Cr^{l+1},\ \ Z(r) = -Cr^l/l$ near the center of the star and $\omega^2 r^2 e^{\Lambda - 2\Phi} Z + \Phi^{\prime}Q\Big|_{r=R}=0$ at the surface of the star. Once we have the EOS of NS matter, we can solve these TOV and pulsating equations to determine the $f$-mode oscillations. In the present study, we attempt to study the effects of axions on dense quark matter EOS and hence on the frquencies of the f-mode oscillations. For this purpose, to describe quark matter interacting with axions, we consider a three flavor  Nambu--Jona-Lasino (NJL) model with the Lagrangian \cite{Sakai:2011gs, Chatterjee:2014csa}
\begin{eqnarray}
    \mathcal{L} &=& \bar{q}(i \gamma^\mu {\partial_\mu} - \hat{m})q + G_s \sum_{A=0}^8\big[ (\bar{q}\lambda^A q)^2 + (\bar{q}i\gamma_5 \lambda^A q)^2 \big] \nonumber \\
    &-& K \big[ e^{i\theta} \text{det}\lbrace \bar{q}(1+\gamma^5) q \rbrace + e^{-i\theta} \text{det} \lbrace\bar{q}(1-\gamma^5) q\rbrace\big] - G_v\left[(\bar q\gamma^\mu q)^2+(\bar q\gamma^\mu\gamma^5 q)^2\right].
\end{eqnarray}
where $q=(q_u,q_d,q_s)^T$ are the quark fields, $\hat{m}$ represents the current quark mass matrix $\text{diag}(m_{u},m_{d},m_{s})$. In the present investigation we consider $m_{u}=m_{d}=m_0$. $\lambda^0 = \sqrt{2/3} ~I_{3\times3}$, here $I_{3\times3}$ is the $3\times 3$ identity matrix in flavor space, $\lambda^A$ with $A=1,2,...,8$ are the Gell-Mann matrices in flavor space. The parameter $G_s$ denotes the coupling of the four-quark interaction which includes scalar and pseudo-scalar type interactions. $K$ is the coupling of the Kobayashi-Maskawa -'t Hooft determinant interaction. For the given Lagrangian, we can find EOS of quark matter at high densities. For the low densities, we consider the Walecka type relativistic mean field (RMF) model as \cite{Mishra:2001py}
\begin{eqnarray}
\mathcal{L} &=& \sum_i \bar{\Psi}_i\left( i\gamma_{\mu}\partial^{\mu} - m_i+g_{\sigma}\sigma - g_{\omega}\gamma_{\mu}\omega^{\mu}- g_{\rho}\gamma_{\mu}\vec{I}_i\vec{\rho}^{\mu}\right)\Psi_i + \mathcal{L}_{\rm{mes}}
\end{eqnarray}
and
\begin{eqnarray}
\mathcal{L}_{\rm mes} &=& \frac{1}{2}\partial_{\mu}\sigma\partial^{\mu}\sigma - \frac{1}{2} m_{\sigma}^2\sigma^2 - \frac{1}{4}\Omega^{\mu \nu}\Omega_{\mu \nu}  + \frac{1}{2}m_{\omega}^2\omega_{\mu}\omega^{\mu} - \frac{1}{4}\vec{R}^{\mu \nu}\vec{R}_{\mu \nu}+\frac{1}{2}m_{\rho}^2\vec{\rho}_{\mu}\vec{\rho}^{\mu} \nonumber \\
&& - \frac{\kappa}{3!}(g_{\sigma {\rm N}}\sigma)^3 + \frac{\lambda}{4!}(g_{\sigma {\rm N}}\sigma)^4 - \frac{\xi}{4!}(g_{\omega {\rm N}}^2\omega_{\mu}\omega^{\mu})^2 - {\Lambda^\prime}(g_{\omega {\rm N}}^2\omega_{\mu}\omega^{\mu})(g_{\rho {\rm N}}^2\rho_{\mu}\rho^{\mu})     
\end{eqnarray}
where, $\Psi_{\rm i}$ and $m_{\rm i}$ correspond to the baryonic field and its bare mass respectively, $g_{\alpha}$ for $\alpha \in \sigma, \omega^{\mu}, {\bm \rho}, \phi^{\mu}$ are the coupling constants of mesons with baryons. The $\Omega^{\mu\nu} = \partial^{\mu}\omega^{\nu} - \partial^{\nu}\omega^{\mu}$, ${\rm \bf R}^{\mu\nu} = \partial^{\mu}{\rm \bm \rho}^{\nu} - \partial^{\nu}{\rm \bm \rho}^{\mu}$ are the vector mesonic field strength tensor. At the mean field level i.e. $\langle \sigma \rangle = \sigma_0$, $\langle \omega_{\mu} \rangle = \omega_0 \delta_{\mu 0}$, and $\langle \rho_{\mu}^{a}\rangle = \delta_{\mu 0}\delta_{3}^a \rho_3^0$, we can redefine the effective baryon mass and chemical potential as,
\begin{eqnarray}
m_i^{*}   &=& m_i - g_{\sigma} \sigma_0 \\
\mu_i^{*} &=& \mu_i - g_{\omega}\omega_0 - g_{\rho} I_{3i}\rho_{3}^0
\end{eqnarray}
As for the quark matter, we can also find EOS for the hadronic matter for the given RMF Lagrangian. Now we use the Gibbs construction mechanism, $p_{\rm HP} (\mu_{\rm B}^c, \mu_{\rm E}^c) = p_{\rm QP} (\mu_{\rm B}^c, \mu_{\rm E}^c) = p_{\rm MP} (\mu_{\rm B}^c, \mu_{\rm E}^c)$ (where HP: hadronic phase, MP: mixed phase, and QP: quark phase. $\mu_{\rm B}^c$ and $\mu_{\rm E}^c$ are the critical baryon and electric chemical potentials respectively.), to get HQPT and a EOS for hybrid stars for various values of axion parameter $\theta$. 

\section{Results and discussions}
\begin{figure}[th]
\centering
\includegraphics[scale=0.38]{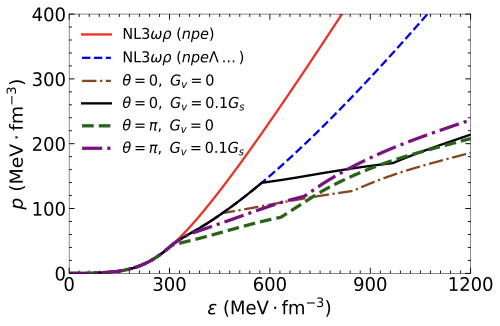}
\includegraphics[scale=0.38]{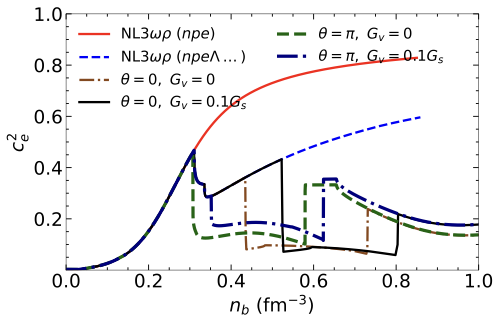}
\caption{Equation of state of neutron/hybrid stars (left) and the variation of the equilibrium speed of sound in the charge neutral matter as a function of baryon number density (right) with axion parameter $\theta$ and vector coupling in the NJL model, $G_v$.}\label{eos_cs2}
\end{figure}
In Figure \ref{eos_cs2}(left), we plot the EOSs for different values of $\theta$'s as displayed with legends. As expected, the inclusion of hyperons softens the hadronic EOS as is clearly seen in the figure. For $\theta = 0,\ G_{v} = 0.0$, MP starts at energy density $\epsilon_1 = 463.7\ {\rm MeV/fm}^{3}$ and ends at $\epsilon_2 = 842.7\ {\rm MeV/fm}^{3}$. The corresponding baryon density for the beginning of MP $n_{\rm b}^{(1)} \sim 0.44\ {\rm fm}^{-3} \equiv 2.75n_0$, ($n_0 = 0.16\ {\rm fm}^{-3}$) and ends at $n_{\rm b}^{(2)} \sim 0.73\ {\rm fm}^{-3} \equiv 4.56 n_0$ beyond which the matter is in a pure quark matter phase. Similarly, for maximum CP violation effect at $\theta = \pi$ at $G_{v} = 0$, the threshold for MP $n_{\rm b}^{(1)}(\theta = \pi) \sim 0.31\ {\rm fm}^{-3} \equiv 1.9n_0$. Increasing the value of $G_v = 0.1 G_s$ makes EOS stiffer resulting the threshold for the appearance of MP to a higher value. 
%
In Figure \ref{eos_cs2}(right), we plot the variation of speed of sound ($c_e^2$) for the charged neutral matter as a function of baryon number density for various cases as displayed in the legend. As the density is increases in HP, $c_e^2$ increases monotonically and saturating at about $c_e^2 \sim 0.8$ at high densities. The maximum value of $c_e^2$ reaches to about 0.5 when the $\Lambda$ hyperons start to appear. For each hyperon, we see a kink type signature in the monotonic curve. For the case of $\theta = \pi$ and $G_v = 0$, MP starts appearing at $n_{\rm B} = 0.31\ {\rm fm}^{-3}$ which is a little below the $\Lambda$ hyperon threshold for the hyperonic matter. The sound speed decreases discontinuously from about $c_e^2 = 0.6$ to $c_e^2 = 0.12$ beyond which it shows a continuous behavior till the end of mixed phase where it again discontinuously increases from $c_e^2 = 0.1$ to $c_e^2 = 0.33$ of pure QP. As the density increases further in the QP at $n_{\rm B} = 0.67\ {\rm fm}^{-3}$ the sound velocity start decreasing due to appearance of strange quark which softens EOS. Similar effects can also be seen for other cases too. 

\begin{figure}[th]
\centering
\includegraphics[scale=0.38]{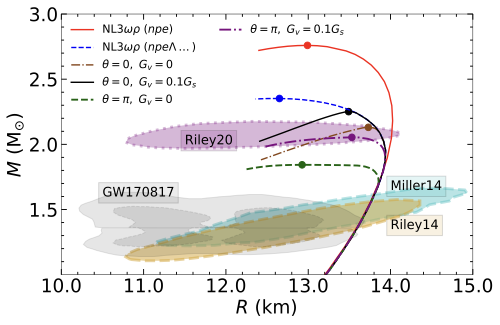}
\includegraphics[scale=0.38]{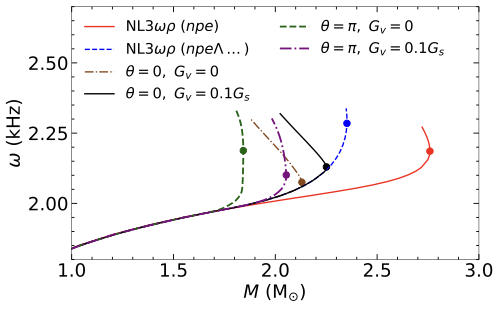}
\caption{Left:Mass-radius relation for various cases. Right: the $f$ mode frequency of different mass NS for various cases of the axion parameter $\theta$ and the vector coupling $G_v$ in the NJL model.}\label{mr_fmode}
\end{figure}
In Figure \ref{mr_fmode} (left), we show the mass-radius curves for the various cases as displayed in the legend of the figure. The dots in each curves represent the maximum mass star of corresponding curve. We see that all the results, except $\theta=\pi$ and $G_v=0$, follow the astrophysical observations. Appearance of hyperons and even quark matter, EOS softens and hence give small mass NSs. However, the vector coupling in the NJL model stiffens the EOS and give larger mass NSs. 
%
In Figure \ref{mr_fmode}(right), we display the $f$ mode frequency of different mass NS for various cases as displayed in the legend. We see that the appearance of hyperons in the matter enhances the $f$ mode frequencies of similar mass NS without hyperons. Further more, appearance of quark matter with axionic effects (non-zero $\theta$) enhances $f$-mode frequencies of a NS more as compared to the same mass NS without hyperons and quark matter in the core. The largest effects can be seen for the case when $\theta=\pi$ and $G_v=0$ but it does not satisfy the maximum mass constraint. So we can say that when $\theta = \pi$ and $G_v=0.1$ gives larger enhancement and satisfies all the astrophysical constraints.

\section{Summary and conclusion}
It is observed that  presence of axions reduces the critical density for chiral transition. Using charge neural and beta equlibriataed quark matter, we discussed the gross structural properties of hybrid neutron stars.This approach allows for stable hybrid NSs containing quark matter in a mixed phase with hyperonic matter, with or without axions, while satisfying modern astrophysical constraints. With vector interactions, it is possible to have  stable hybrid NSs with a quark matter core, where the quark matter can exist either in a pure phase or in a mixed phase with hyperonic matter, surrounded by an outer core of hyperonic matter while satisfying maximum mass constraints. 
 It is observed that the f mode frequncies of hybrid stars  get enhanced
due to presence of hyperons as well as quark matter as compared to nucleonic matter.The enhancement is particularly large in the presence of axions.
Detection of the enhanced f mode frequency could be indicative of non-nucleonic degrees of freedom in neutron star matter.


\bibliographystyle{elsarticle-num-names} 
\bibliography{bib}

\end{document}